\documentstyle[fleqn,twoside,amsfonts]{article}

\makeatletter

\def\noi{\noindent}

\renewcommand{\section}{\@startsection{section}{1}{0pt}%
        {-3.5ex plus -1ex minus -.2ex}{2.3ex plus .2ex}%
        {\large\bf\protect\raggedright}}

\renewcommand{\subsection}{\@startsection{subsection}{2}{0pt}%
        {-3ex plus -1ex minus -.2ex}{1.4ex plus .2ex}%
        {\normalsize\bf\protect\raggedright}}

\renewcommand{\thesubsubsection}%
        {\arabic{section}.\arabic{subsection}.\arabic{subsubsection}.}

\renewcommand{\@oddhead}{\raisebox{0pt}[\headheight][0pt]{%
   \vbox{\hbox to\textwidth{\rightmark \hfil \rm \thepage \strut}\hrule}}}
\renewcommand{\@evenhead}{\raisebox{0pt}[\headheight][0pt]{%
   \vbox{\hbox to\textwidth{\thepage \hfil \leftmark \strut}\hrule}}}
\newcommand{\heads}[2]{\markboth{\protect\small\it #1}{\protect\small\it
#2}}
\newcommand{\Acknow}[1]{\subsection*{Acknowledgement} #1}

\newcommand{\Title}[1]{\noi {\Large #1} \\}

\newcommand{\Abstract}[1]{\vskip 2mm \begin{center}
        \parbox{16.4cm}{\small\noi #1} \end{center}\medskip}

\newcommand{\foom}[1]{\protect\footnotemark[#1]}

\newcommand{\email}[2]{\footnotetext[#1]{e-mail: #2}}
\newcommand{\eps}{ \varepsilon }

\newcommand{\sect}[1]{Sec.\,#1}
\def\nq{\hspace*{-1em}}
\def\nqq{\hspace*{-2em}}
\def\nhq{\hspace*{-0.5em}}

\def\cm{\hspace*{1cm}}
\def\inch{\hspace*{1in}}

\def\al{&\nhq}
\def\lal{&&\nqq {}}

\def\eqs{Eqs.\,}
\def\beq{\begin{equation}}
\def\eeq{\end{equation}}
\def\bear{\begin{eqnarray}}
\def\bearr{\begin{eqnarray} \lal}
\def\ear{\end{eqnarray}}
\def\earn{\nonumber \end{eqnarray}}
\def\nn{\nonumber\\ {}}

\def\nnn{\nonumber\\ \lal }

\def\eql{\al =\al}

\def\e{{\,\rm e}}
\def\d{\partial}

\def\im{\mathop{\rm Im}\nolimits}

\def\sign{\mathop{\rm sign}\nolimits}

\def\dim{\mathop{\rm dim}\nolimits}
\def\const{{\rm const}}

\makeatother

\def\beq#1{\begin{equation}\label{#1}}
\def\eeq{\end{equation}}
\def\ber#1{\begin{eqnarray}\label{#1} \nqq}
\def\eer{\end{eqnarray}}
\renewcommand{\nn}{\nonumber}
\renewcommand{\bear}[1]{\begin{eqnarray}\label{#1}}
\newcommand{\Authorss}[5]{\noi
        {\large\bf #1\dag\ #2\ddag\ #3\dag}
 \medskip\begin{description}
        \item[\dag]{\it #4} \item[\ddag]{\it #5}
 \end{description}}

\catcode`\@=11 \@addtoreset{equation}{section}\catcode`\@=12

\newcommand{\N}{{\mathbb N}}
\newcommand{\R}{{\mathbb R}}
\renewcommand{\im}{i}
\newcommand{\p}{\partial}
\newcommand{\btd}{\bigtriangledown}
\newcommand{\btu}{\bigtriangleup}
\newcommand{\tri}{\Delta}

\newcommand{\ch}{\cosh}

\heads{V.D. Ivashchuk, M. Kenmoku and V.N. Melnikov}
      {On quantum analogues of p-brane black hole}

\begin{document}
\twocolumn[

\thispagestyle{empty}

\Title{ON QUANTUM ANALOGUES OF P-BRANE BLACK HOLES}

\Authorss{V.D. Ivashchuk\foom 1,} {M. Kenmoku\foom 2}
  {and V.N. Melnikov\foom 3}
         {Center for Gravitation and Fundamental Metrology,
         VNIIMS, 3-1 M. Ulyanovoy Str., Moscow 117313, Russia \\
         Institute of Gravitation and Cosmology, PFUR,
         6 Miklukho-Maklaya St., Moscow 117198, Russia}
         {Department of Physics, Nara Women University, Nara 630, Japan}

\Abstract
{In a multidimensional model with several scalar fields
and an $m$-form we deal with classical spherically symmetric solutions
with one (electric or magnetic) $p$-brane and Ricci-flat internal spaces
and the corresponding solutions to the Wheeler--DeWitt (WDW) equation.
Classical black holes are considered and their quantum analogues (e.g. for
$M2$ and $M5$ extremal solutions in $D =11$ supergravity, electric and
magnetic charges in $D=4$ gravity) are suggested when the curvature
coupling in the WDW equation is zero.}

] 

\email 1 {ivas@rgs.phys.msu.su}
\email 2 {kenmoku@phys.nara-wu.ac.jp}
\email 3 {melnikov@rgs.phys.msu.su}

\section{Introduction}

In this paper we continue our investigations of $p$-brane solutions
(see, e.g., \cite{DKL,St,AIR} and references therein)
based on the sigma-model approach  \cite{IM4,IM,IMR}.

The model under consideration contains several
scalar dilatonic fields and one antisymmetric form.
We consider spherically symmetric solutions (see \cite{BIM,IMJ}),
when all functions depend on one radial variable and pay attention
to black hole (BH) solutions with Ricci-flat
internal spaces (see \cite{BIM,IMJ} and special solutions
in \cite{CT,AIV,O}).

The corresponding solutions to the WDW equation (in the
spherically symmetric case) were considered in \cite{GrIM,IMJ}.  Here we
use the covariant ``d'Alembertian" (and/or conformally-covariant) form of
the WDW equation of Refs.\,\cite{Mis,Hal,IMZ}.  We single
out certain classes of solutions to the WDW equations and suggest quantum
analogues of BH solutions.

The plan of the paper is as follows.
In \sect  2 we consider the model and present the WDW equation.
\sect  3 is devoted to classical and quantum exact solutions
when spherical symmetry is assumed. In \sect 4 we consider
the classical black-hole solutions and for the case $a=0$
(the term $a R[{\cal G}]$ in the WDW equation is responsible for
the scalar curvature of minisupermetric ${\cal G}$) we suggest quantum
analogues of black-hole solutions. In the extremal case we
consider several examples, e.g. for $D=4$ Einstein-Maxwell
theory and $D=11$ supergravity. It is shown that the  brane part of the
solution satisfying the outgoing-wave boundary condition is regular for
small  brane quasi-volume (it looks like a pseudo-Euclidean quantum
wormhole). We also compare our approach with that suggested by H. L\"u,
J. Maharana, S. Mukherji  and C.N. Pope \cite{LMMP} (with flat
minisupermetric  and classical fields of forms).

\section{The model}

Consider the model governed by the action
\bear{2.1i}\lal
    S = \frac{1}{2\kappa^{2}}
 \int_{M} d^{D}z \sqrt{|g|} \{ {R}[g] - h_{\alpha\beta}
  g^{MN} \d_{M} \varphi^\alpha \d_{N} \varphi^\beta
   \nnn \cm
 - \frac{1}{m!} \exp[ 2 \lambda (\varphi) ] F^2 \}+ S_{GH}
\ear
where $g = g_{MN} dz^{M} \otimes dz^{N}$ is the metric
($M,N =1, \ldots, D$), $\varphi=(\varphi^\alpha)\in \R^l$
is a vector of dilatonic scalar fields,
$(h_{\alpha\beta})$ is a non-degenerate $l\times l$ matrix ($l\in \N$),
$F =  dA = \frac{1}{m!} F_{M_1 \ldots M_{m }}
dz^{M_1} \wedge \ldots \wedge dz^{M_{m}}$
is an $m$-form ($m   \geq 1$) on a $D$-dimensional manifold $M$
and $\lambda$ is a $1$-form
on $\R^l$: $\lambda (\varphi) =\lambda_{ \alpha} \varphi^\alpha$,
$\alpha=1,\ldots,l$. In (\ref{2.1i}) we denote
$|g| = |\det (g_{MN})|$, $F^2 =
F_{M_1 \ldots M_{m}} F_{N_1 \ldots N_{m}}
g^{M_1 N_1} \ldots g^{M_{m} N_{m  }}$, where $S_{\rm GH}$ is the
standard Gibbons-Hawking boundary term \cite{GH}.
The signature of the metric is $(-1,+1, \ldots, +1)$.

The equations of motion corresponding to  (\ref{2.1i}) have the following
form:
\bear{2.4i}
 R_{MN} - \frac{1}{2} g_{MN} R  \eql  T_{MN},       \\
\label{2.5i}
 {\btu}[g] \varphi^\alpha -
 \sum_{a \in \Delta}   \frac{\lambda^{\alpha}}{m  !}
  \e^{2 \lambda(\varphi)} F^2 \eql 0,         \\
\label{2.6i}
 \nabla_{M_1}[g] (\e^{2 \lambda(\varphi)}
    F^{M_1 \ldots M_{m }})  \eql  0,
\ear
$\alpha=1,\ldots,l$.
In (\ref{2.5i}) $\lambda^{\alpha} = h^{\alpha \beta}
\lambda_{ \beta }$, where $(h^{\alpha \beta})$
is matrix inverse to $(h_{\alpha \beta})$. In (\ref{2.4i})
\bear{2.7i}
 T_{MN} =   T_{MN}[\varphi,g]+  \e^{2 \lambda(\varphi)} T_{MN}[F,g],
\ear
where
\bear{2.7ii}
 && \nqq T_{MN}[\varphi,g] =
 h_{\alpha\beta}\left(\p_{M} \varphi^\alpha \p_{N} \varphi^\beta -
 \frac{g_{MN}}2 \p_{P} \varphi^\alpha \p^{P} \varphi^\beta\right),
\nnn
 T_{MN}[F,g] = \frac{1}{m!}\left[ - \frac{1}{2} g_{MN} F^{2}
\nn \right. \\
&& \left.
 + m   F^{a}_{M M_2 \ldots M_{m }} F_{N}^{ M_2
     \ldots M_{m }}\right].
\earn
In (\ref{2.5i}), (\ref{2.6i}) ${\btu}[g]$ and ${\btd}[g]$
are the Laplace-Beltrami and covariant derivative operators
corresponding to  $g$, respectively.

Consider the manifold
\beq{2.10g} \nhq
M = \R  \times (M_0 = S^{d_0}) \times (M_1 = \R) \times \ldots \times M_{n}
\eeq
with the metric
\beq{2.11g}
g= \e^{2{\gamma}(u)} du \otimes du + \sum_{i=0}^{n} \e^{2\phi^i(u)} g^i ,
\eeq
where $u$ is a radial coordinate,
$g^i  = g^i_{m_{i} n_{i}}(y_i) dy_i^{m_{i}} \otimes dy_i^{n_{i}}$
is a metric on $M_{i}$  satisfying the equation
\beq{2.12g}
 R_{m_{i}n_{i}}[g^i] = \xi_{i} g^i_{m_{i}n_{i}},
\eeq
$m_{i},n_{i}=1,\ldots,d_{i}$; $d_{i} = \dim M_i$, $\xi_i= \const$,
$i=0,\dots,n$; $n \in \N$. Thus $(M_i,g^i)$ are Einstein spaces.
The functions $\gamma,\phi^i$: $(u_-,u_+)\to\R$ are smooth.
The metric $g^0$ is a canonical metric on $M_0 = S^{d_0}$, $\xi_0 = d_0 -1$
and $g^1 = - dt \times dt$, $\xi_1 = 0$. Here  $(M_1,g^1)$
is a time manifold.

Each manifold $M_i$ is assumed to be oriented and connected,
$i = 0,\ldots,n$ (for $i=0,1$ this is satisfied automatically). Then the
volume $d_i$-form
\beq{2.13g} \tau_i  = \sqrt{|g^i(y_i)|} \ dy_i^{1} \wedge
\ldots \wedge dy_i^{d_i},
\eeq
and the signature parameter
\beq{2.14g}
\eps(i)  = \sign \det (g^i_{m_{i}n_{i}}) = \pm 1 \eeq are correctly defined
for all $i=0,\ldots,n$. Here $\eps(0) = 1$ and $\eps(1) = -1$.

Let $\Omega_0$ be a set of all subsets of $I_0\equiv\{ 0, \ldots, n \}$:
$\Omega_0= \{ \emptyset, \{ 0 \}, \{ 1 \}, \ldots, \{ n \},
\{ 0, 1 \}, \ldots, \{ 0, 1,  \ldots, $ $n \} \}$
For any $I = \{ i_1, \ldots, i_k \} \in \Omega_0$,
$i_1 < \ldots < i_k$,   we define the form
\beq{2.17i}
 \tau(I) \equiv \tau_{i_1}  \wedge \ldots \wedge \tau_{i_k},
\eeq
of rank
\beq{2.19i}
 d(I) \equiv  \sum_{i \in I} d_i,
\eeq
and the corresponding $p$-brane submanifold
\beq{2.18i}
 M_{I} \equiv M_{i_1}  \times  \ldots \times M_{i_k},
\eeq
where $p=d(I)-1$, $\dim M_{I} = d(I)$. We also define the $\eps$-symbol
\beq{2.19e}
 \eps(I) \equiv  \eps(i_1) \ldots \eps(i_k).
\eeq
For $I = \emptyset$ we put  $\tau(\emptyset) = \eps(\emptyset) = 1$,
$d(\emptyset) = 0$.

For the field of form we adopt the following 1-form
ansatz
\beq{2.27n}
 F= {\cal F}^s,
\eeq
where
\bear{2.28n}
      {\cal F}^s \eql d\Phi^s \wedge\tau(I_s), \inch s = e,\\ \label{2.29n}
 {\cal F}^s \eql \e^{-2\lambda (\varphi)}
  *\left(d\Phi^s \wedge\tau(I)\right),\ \qquad s = m
\ear
and $I_s \in\Omega$.
In (\ref{2.29n}) $*=*[g]$ is the Hodge operator on $(M,g)$.
The indices $e$ and $m$ correspond to electric and magnetic
$p$-branes, respectively.

For the potentials in (\ref{2.28n}), (\ref{2.29n})
and the dilatonic scalar fields we put
\beq{2.28nn}
 \Phi^s=\Phi^s(u), \qquad \varphi^\alpha=\varphi^\alpha(u),
\eeq
$s = e,m$; $\alpha=1,\dots,l$.

>From (\ref{2.28n})  and (\ref{2.29n}) we obtain
the relations between the dimensions of the $p$-brane
worldsheets and the ranks of forms:
\bear{2.d1}
 d(I_s) = m   - 1,  \quad d(I_s) = D - m   - 1,
\ear
for $s = e,m$, respectively.

It follows from \cite{IM} that the
equations of motion (\ref{2.4i})--(\ref{2.6i}) and the Bianchi identities
\beq{2.b}
 dF=0,
\eeq
for the field configuration (\ref{2.11g}), (\ref{2.27n})--(\ref{2.28nn})
are equivalent to equations of motion for a $\sigma$-model
with the action
\bear{2.25gn} \lal
 S_{\sigma} = \frac{\theta}2
 \int du \,{\cal N} \biggl\{G_{ij}\dot\phi^i\dot\phi^j
 +h_{\alpha\beta}\dot\varphi^{\alpha}\dot\varphi^{\beta}
\nnn
 + \eps_s\exp[-2U^s(\phi,\varphi)](\dot\Phi^s)^2
   -2{\cal N}^{-2}V(\phi)\biggr\}
\ear
where $\dot x\equiv dx/du$,
\beq{2.27gn}
 V = {V}(\phi) =
\frac{1}{2} \sum_{i =0}^{n} \xi_i d_i \e^{-2 \phi^i + 2 {\gamma_0}(\phi)}
\eeq
is the potential with
\beq{2.24gn}
 \gamma_0(\phi)= \sum_{i=0}^nd_i\phi^i;  \label{2.32g}
\eeq
furthermore,
\beq{2.24gn1}
 {\cal N}=\exp(\gamma_0-\gamma)>0
\eeq
is the lapse function,
\bear{2.u}
 U^s \eql U^s(\phi,\varphi)= -\chi_s\lambda(\varphi) +
   \sum_{i\in I_s}d_i\phi^i, \\ \label{2.e}
\eps_s \eql (-\eps[g])^{(1-\chi_s)/2}\eps(I_s)\theta = \pm 1
\ear
for $s= e, m$, $\eps[g]= \sign \det (g_{MN})$,
$\chi_s=+1, -1$, for $s= e,m$, respectively,
and
\beq{2.c}
G_{ij}=d_i\delta_{ij}-d_id_j
\eeq
are components of the ``cosmological'' minisupermetric, $i,j=0,\dots,n$
\cite{IMZ}.

In the electric case  for finite ``internal space''
volumes $V_i$ the action (\ref{2.25gn}) (the time manifold should be $S^1$)
coincides with the action (\ref{2.1i}) if
$\theta=-1/\kappa_0^2$, $\kappa^{2} = \kappa^{2}_0 V_0 \ldots V_n$.

The action (\ref{2.25gn}) may be also written in the form
\beq{2.31n}                             \nhq
 S_\sigma=\frac\theta2\int du{\cal N}\left\{
 {\cal G}_{\hat A\hat B}(X)\dot X^{\hat A}\dot X^{\hat B}-
  2{\cal N}^{-2}V \right\}
\eeq
where $X = (X^{\hat A})=(\phi^i,\varphi^\alpha,\Phi^s)\in{\R}^{N}$, and the
minisupermetric
${\cal G}={\cal G}_{\hat A\hat B}(X)dX^{\hat A}\otimes dX^{\hat B}$
on minisuperspace ${\cal M}={\R}^{N}$,
$N = n+ 2 +l$, is defined by the relation
\beq{2.33n}
 ({\cal G}_{\hat A\hat B}(X))=\left(\begin{array}{ccc}
 G_{ij}&0&0\\[5pt]
 0&h_{\alpha\beta}&0\\[5pt]
 0&0&\eps_s\e^{-2U^s(X)}
  \end{array}\right).
\eeq

The minisuperspace metric may be written as follows:
\beq{2.34n}
 {\cal G}=\bar G+ \eps_s\e^{-2U^s(x)}d\Phi^s\otimes d\Phi^s
\eeq
where $x=(x^A)=(\phi^i,\varphi^\alpha)$,
$\bar G=\bar G_{AB}dx^A\otimes dx^B$,
\bear{2.36n}
(\bar G_{AB})=\left(\begin{array}{cc}
G_{ij}&0\\
0&h_{\alpha\beta}
\end{array}\right),
\ear
 $U^s(x) = U_A^sx^A$ is defined in  (\ref{2.u}) and
\beq{2.38n}
 (U_A^s)=(d_i\delta_{iI_s},-\chi_s\lambda_{\alpha}).
\eeq
Here $\delta_{iI}$ is an indicator of $i$ belonging to $I$:
$\delta_{iI} = 1$, $i \in I$ and $\delta_{iI} = 0$, $i \notin I$.

The potential (\ref{2.27gn}) reads
\beq{2.40n}
V=\sum_{j=0}^n \frac{\xi_j}{2} d_j \e^{2U^j(x)},
\eeq
where
\bear{2.41n}
 U^j(x) \eql U_A^jx^A=-\phi^j+\gamma_0(\phi), \\ \label{2.43n}
 (U_A^j) \eql (-\delta_i^j+d_i,0).
\ear

The integrability of the Lagrange system (\ref{2.31n}) depends
on the scalar products of co-vectors $U^j$, $U^s$ corresponding to $\bar
G$:
\beq{2.45n}
 (U,U')=\bar G^{AB}U_AU'_B,
\eeq
where
\beq{2.46n}
 (\bar G^{AB})=\left(\begin{array}{cc} G^{ij}&0\\
    0&h^{\alpha\beta}
 \end{array}\right)
\eeq
is the matrix inverse to (\ref{2.36n}). Here (as in \cite{IMZ})
\beq{2.47n}
 G^{ij}=\frac{\delta^{ij}}{d_i}+\frac1{2-D},
\eeq
$i,j=0,\dots,n$. These products have the following form:
\bear{2.48n}
 (U^i,U^j) \eql \frac{\delta_{ij}}{d_j}-1, \\ \label{2.51n}
 (U^s,U^{s}) \eql d(I_s)+\frac{(d(I_s))^2}{2-D}
  +\lambda_{\alpha}\lambda_{\beta} h^{\alpha\beta},
    \\ \label{2.52n}
 (U^s,U^i) \eql -\delta_{iI_s},
\ear
$s= e, m$.

\subsection{Wheeler--DeWitt equation}

Here we fix the gauge as follows:
\beq{4.1n}
 \gamma_0-\gamma=f(X),  \quad  {\cal N} = \e^f,
\eeq
where $f$: ${\cal M}\to{\R}$ is a smooth function. Then we obtain the
Lagrange system with the Lagrangian
\beq{3.14r}
 L_f=\frac\theta2\e^f{\cal G}_{\hat A\hat B}(X)
   \dot X^{\hat A}\dot X^{\hat B}-\theta\e^{-f}V
\eeq
and the energy constraint
\beq{3.15r}
 E_f=\frac\theta2\e^f{\cal G}_{\hat A\hat B}(X)
  \dot X^{\hat A}\dot X^{\hat B}+\theta\e^{-f}V=0.
\eeq

The standard prescriptions of covariant and conformally covariant
quantization (see, e.g., \cite{IMZ,Mis,Hal}) lead to the Wheeler-DeWitt
(WDW) equation
\bear{4.2n} \lal \nhq
 \hat{H}^f \Psi^f \equiv
  (-\frac{1}{2\theta}\Delta\left[\e^f{\cal G}\right]+
  \frac{a}{\theta}R\left[\e^f{\cal G}\right]+ \e^{-f} \theta V) \Psi^f=0
\nnn
\ear
where
\beq{4.3n}
 a=a_c(N)=\frac{N-2}{8(N-1)}.
\eeq
Here $\Psi^f=\Psi^f(X)$ is the wave function
corresponding to the $f$-gauge (\ref{4.1n}) and satisfying the relation
\beq{4.3an}
 \Psi^f= \e^{bf} \Psi^{f=0}, \quad b = (2-N)/2,
\eeq
$\Delta[{\cal G}_1]$ and $R[{\cal G}_1]$ denote the Laplace-Beltrami
operator and the scalar curvature corresponding to ${\cal G}_1$. We note
that parameter $a$ may be arbitrary if we do not care about the conformal
covariance of the WDW equation.

For the scalar curvature of minisupermetric (\ref{2.34n})
we get (see (2.29) in \cite{IM}):
\beq{4.4n}
 R[{\cal G}]=- 2(U^s,U^s).
\eeq

For the Laplace operator we obtain
\bear{4.5n} \lal
 \tri[{\cal G}] =\e^{U^s(x)}\frac\d{\d x^A}\left(\bar G^{AB}
  \e^{-U^s(x)}\frac\d{\d x^B}\right)
\nnn  \inch
 + \eps_s\e^{2U^s(x)} \left(\frac\d{\d\Phi^s}\right)^2.
\ear

The WDW equation (\ref{4.2n}) for $f=0$ (in the harmonic time gauge)
\beq{4.7n}
 \hat H\Psi = \left(-\frac{1}{2\theta}\Delta[{\cal G}]
 +\frac{a}{\theta}R[{\cal G}]+\theta V\right)\Psi=0,
\eeq
may be rewritten, using \eqs (\ref{4.4n}), (\ref{4.5n}) and
\beq{4.8n} \nq
 U^{si}= G^{ij}U_j^s= \delta_{iI_s}-\frac{d(I_s)}{D-2}, \
  U^{s\alpha}= - \chi_s \lambda^\alpha,
\eeq
as follows:
\bear{4.10n}\lal
 2\theta\hat H\Psi= \biggl\{ -G^{ij}\frac{\d}{\d \phi^i}
 \frac\d{\d\phi^j}-h^{\alpha\beta}\frac{\d}{\d\varphi^{\alpha}}
   \frac{\d}{\d \varphi^{\beta}}
\nnn\cm
  - \eps_s \e^{2U^s(\phi,\varphi)} \left(\frac{\d}{\d \Phi^s} \right)^2
   \nnn
 +\biggl[\sum_{i\in I_s}\frac\d{\d\phi^i}-
 \frac{d(I_s)}{D-2}\sum_{j=0}^n\frac\d{\d\phi^j}-
 \chi_s\lambda_{a_s}^\alpha\frac\d{\d\varphi^\alpha}\biggr]
\nnn  \inch
 + 2aR[{\cal G}]+2\theta^2V \biggr\} \Psi=0.
\ear
Here $\hat H\equiv\hat H^{f=0}$ and $\Psi\equiv\Psi^{f=0}$.

\section{Exact solutions}

Here we use the following restriction on the parameters of the model:
\beq{5.1n}\nq
 {\bf (i)} \qquad  \xi_0 = d_0 -1, \qquad    \xi_1=\ldots=\xi_n=0,
\eeq
one space $M_0 = S^{d_0}$ is a unit sphere and
all $M_i$ ($i > 1$) are Ricci-flat;
\beq{5.3n}\nq
 {\bf (ii)} \cm 0\notin I_s,
\eeq
i.e. the ``brane'' submanifold $M_{I_s}$ (see (\ref{2.18i})) does not
contain $M_0$, and
\beq{5.5n}  \nq
 {\bf (iii)}\qquad (U^s,U^s) > 0.
\eeq
The latter is satisfied in most of examples of interest.

>From {\bf (i)}, {\bf (ii)}  we get for the potential (\ref{2.40n}):
\beq{5.6n}
 V=\frac12 \xi_0 d_0\e^{2U^0(x)},
\eeq
where (see (\ref{2.48n}))
\beq{5.7n}
 (U^0,U^0)= 1/d_0 -1 < 0.
\eeq

>From {\bf (iii)} and (\ref{2.52n}) we get
\beq{5.8n}
 (U^0,U^{s})=0.
\eeq

\subsection{Classical solutions}

Consider a solution to the Lagrange equations corresponding to the
Lagrangian (\ref{3.14r}) with the energy-constraint (\ref{3.15r})
under the restrictions (\ref{5.1n})--(\ref{5.5n}). We put
$f=0$, i.e., use the harmonic time gauge.

Integrating the Maxwell
equations (for $s = e$) and the Bianchi identities (for $s = m$), we get
\bear{5.29n}
 \frac d{du}\left(\exp(-2U^s)\dot\Phi^s\right)=0
 \Longleftrightarrow
  \dot\Phi^s=Q_s \exp(2U^s),
\nnn
\ear
where $Q_s=\const$. We put $Q_s\ne 0$.

For fixed $Q_s$, the Lagrange equations for the Lagrangian (\ref{3.14r})
with $f=0$ corresponding to $(x^A)=(\phi^i,\varphi^\alpha)$, with \eqs
(\ref{5.29n}) substituted, are equivalent to the Lagrange equations for
the Lagrangian
\beq{5.31n}
 L_Q=\frac12\bar G_{AB}\dot x^A\dot x^B-V_Q
\eeq
where
\beq{5.32n}
 V_Q=V+\frac12 \eps_s Q_s^2\exp[2U^s(x)],
\eeq
$(\bar G_{AB})$ and $V$ are defined in (\ref{2.36n}) and (\ref{2.27gn}),
respectively. The zero-energy constraint (\ref{3.15r}) reads
\beq{5.33n}
 E_Q=\frac12\bar G_{AB}\dot x^A\dot x^B+V_Q=0.
\eeq

When the conditions {\bf (i)}--{\bf (iii)} are satisfied, exact solutions
for the Lagrangian (\ref{5.31n}) with the potential (\ref{5.32n}) and $V$
>from (\ref{5.6n}) have the following form \cite{IMJ}:
\bear{5.34n} \lal
 x^A(u)=-\frac{U^{0A}}{(U^0,U^0)}\ln |f_0(u-u_0)|
 \nnn
 - \frac{U^{sA}}{(U^s,U^s)}\ln |f_s(u-u_s)| + c^A u + \bar{c}^A
\ear
where $u_0$, $u_s$ are constants,
\bear{5.35n}
 f_0(\tau)= \frac{(d_0-1)}{\sqrt{C_0}} \sinh(\sqrt{C_0}\tau),
\ear
and
\bear{5.39n}  \lal
 f_s(\tau)=\frac{|Q_s|}{\nu_s \sqrt{C_s}}\sinh(\sqrt{C_s}\tau),
   \qquad \eps_s<0; \\ \label{5.41n}
   \lal \cm
\frac{|Q_s|}{\nu_s\sqrt{C_s}}\ch(\sqrt{C_s}\tau), \quad
  C_s>0, \quad \eps_s>0;
\ear
$C_0$ and $C_s$ are constants. Here $\sinh(\sqrt{C}x)/ \sqrt{C} = x$ for $C
= 0$.

The contravariant components $U^{0A}= \bar G^{AB} U^0_B$ are
\beq{5.43n}
 U^{0i}=-\frac{\delta_0^i}{d_0}, \quad  U^{0\alpha}=0.
\eeq
(For $U^{sA}, s \in S$, see (\ref{4.8n})).

The vectors $c=(c^A)$ and $\bar c=(\bar c^A)$ satisfy the linear constraint
relations (due the configuration space splitting into a sum of three
mutually orthogonal subspaces, see \cite{IMJ})
\bear{5.47n} \lal
 U^0(c)= c^0+\sum_{j=0}^nd_jc^j=0, \quad U^0(\bar c)=0,
\\ \label{5.49n} \lal
 U^s(c)= \sum_{i\in I_s}d_ic^i-\chi_s\lambda_{a_s\alpha}c^\alpha=0,
 \ U^s(\bar c)= 0.
\ear

The zero-energy constraint
$E=E_0+ E_s+ (1/2)$ $\times\bar G_{AB}c^Ac^B=0$,
with $C_0=2E_0(U^0,U^0)$, $C_s=2E_s(U^s,U^s)$ may be written as
\bear{5.55n}  \lal
 C_0\frac{d_0}{d_0-1}= C_s\nu_s^2 +
  h_{\alpha\beta}c^\alpha c^\beta+\sum_{i=1}^nd_i(c^i)^2
 \nnn \inch
 + \frac1{d_0-1}\left(\sum_{i=1}^nd_ic^i\right)^2.
\ear

The following expressions for the metric and scalar fields follows from
(\ref{4.8n} (\ref{5.43n})) and  (\ref{5.43n}):
\bear{5.63n} \lal
 g= [f_s^2(u-u_s)]^{ d(I_s)\nu_s^2/(D-2)} \times
     \nnn \nq
{\times} \biggl\{[f_0^2(u{-}u_0)]^{d_0/(1-d_0)}\e^{2c^0u+2\bar c^0}
  [du {\otimes} du + f_0^2(u{-}u_0)g^0]
     \nnn
 + \sum_{i \ne 0}
[f_s^2(u-u_s)]^{-\nu_s^2 \delta_{i I_s}}\e^{2c^iu+2\bar c^i}g^i\biggr\},
 \\ \lal \label{5.46n}
   \varphi^\alpha= \nu_s^2\chi_s\lambda_{a_s}^\alpha
    \ln |f_s|+c^\alpha u+\bar c^\alpha.
\ear
>From the relation $\exp(2 U^s)=f_s^{-2}$
(following from (\ref{5.8n}), (\ref{5.34n}) (\ref{5.49n})) we get for
the forms:
\bear{5.57n} \nq
 {\cal F}^s \eql Q_s f_s^{-2}du\wedge\tau(I_s),  \\
\label{5.58n} \nq
 {\cal F}^s \eql \e^{-2 \lambda(\varphi)}
  *\left[Q_s f_s^{-2} du \wedge\tau(I_s)\right] =\bar Q_s \tau(\bar I_s)
\ear
for $s =e,m$, respectively, where  $\bar Q_s=Q_s\eps(I_s)\mu(I_s)$
and $\mu(I) =\pm1$ is defined by the relation
$\mu(I) du \wedge \tau(I_0)=\tau(\bar I)\wedge du\wedge\tau(I)$
\cite{IMJ}.

Thus  we obtain exact spherically symmetric solutions  with
internal Ricci-flat spaces  $(M_i,g^i)$, $i=2,\dots,n$ in the
presence of several scalar fields and one form. The solution is presented
by the relations  (\ref{5.46n}), (\ref{5.57n})--(\ref{5.63n})
with the functions $f_0$, $f_s$  defined in (\ref{5.35n})--(\ref{5.41n})
and the relations (\ref{5.47n})--(\ref{5.49n}), (\ref{5.55n})
on the solution parameters $c^A$, $\bar c^A$ $(A=i,\alpha)$, $C_0$,
$C_s$, $\nu_s$.

This solution describes a  charged $p$-brane (electric or magnetic)
``living'' on the submanifold $M_{I_s}$ (\ref{2.18i}), where the set $I_s$
does not contain $0$, i.e. the $p$-brane lives only in the ``internal"
Ricci-flat spaces.

In the non-composite case with several intersecting $p$-branes, solutions
of this type were considered in \cite{GrIM,BGIM} (the electric case) and
\cite{BIM} (the electro-magnetic case). For the composite case see
\cite{IMJ}.

\subsection{Quantum solutions}

The truncated minisuperspace metric (\ref{2.36n}) may be diagonalized
by the linear transformation
\beq{5.12n}
 z^A=S^A{}_Bx^B, \quad (z^A)=(z^0,z^a,z^s)
\eeq
as follows:
\beq{5.13n}\nq
 \bar G=-dz^0\otimes dz^0+
  \eta_sdz^s\otimes dz^s+dz^a\otimes dz^b\eta_{ab},
\eeq
where $a,b=1,\dots,n$; $\eta_{ab} =\eta_{aa} \delta_{ab};
\eta_{aa}= \pm 1$, and
\bear{5.14n}
 q_0z^0=U^0(x), \cm q_s z^s=U^s(x),
\ear
with $q_0 =|(U^0,U^0)|^{1/2}= \left[1- 1/d_0 \right]^{1/2} >0$,
$q_s=\nu_s^{-1}= |(U^s,U^s)|^{1/2}$.

>From (\ref{4.5n}), (\ref{5.12n}), (\ref{5.13n}) we get
\bear{5.18n} \lal
  \tri[{\cal G}]=-\left(\frac\d{\d z^0}\right)^2+
 \eta^{ab}\frac\d{\d z^a}\frac\d{\d z^b}
 \nnn \nq
 + \e^{q_sz^s}\frac\d{\d z^s} \left(\e^{-q_sz^s}\frac\d{\d z^s}\right)
 + \eps_s \e^{2q_sz^s}\left(\frac\d{\d\Phi^s}\right)^2.
\ear

As usual, we seek a solution to the WDW equation
(\ref{4.7n}) by separation of variables, i.e., we put
\beq{5.20n}
 \Psi_*(z)=\Psi_0(z^0) \Psi_s(z^s)\e^{\im P_s\Phi^s}\e^{\im p_az^a}.
\eeq
It follows from (\ref{5.18n}) that $\Psi_*(z)$ satisfies the WDW equation
(\ref{4.7n}) if
\bear{5.21n}
 2\hat H_0\Psi_0\equiv\left\{\left(\frac\d{\d z^0}\right)^2
  +\theta^2 \xi_0d_0\e^{2q_0z^0}\right\}\Psi_0
  \nn \\  = 2{\cal E}_0\Psi_0;
    \\ \label{5.22n}
 2\hat H_s\Psi_s\equiv \biggl\{-\e^{q_sz^s}\frac\d{\d z^s}
   \left(\e^{-q_sz^s}\frac\d{\d z^s}\right)
  \nn \\
 + \eps_sP_s^2\e^{2q_sz^s}\biggr\} \Psi_s=2{\cal E}_s\Psi_s,
\ear
and
\beq{5.23n}
 2{\cal E}_0+\eta^{ab}p_ap_b+2{\cal E}_s+ 2aR[{\cal G}]=0,
\eeq
with $a$ and $R[{\cal G}]$ from (\ref{4.3n}) and (\ref{4.4n}),
respectively.

Linearly independent solutions to \eqs (\ref{5.21n}) and (\ref{5.22n})
have the following form:
\bear{5.24n}
 \Psi_0(z^0) \eql B_{\omega_0}^0
 \left(\sqrt{- \theta^2 (d_0-1) d_0}\frac{\e^{q_0z^0}}{q_0}\right),\\
\label{5.25n}
 \Psi_s(z^s) \eql \e^{q_sz^s/2}B_{\omega_s}^s
  \left(\sqrt{\eps_sP_s^2}\frac{\e^{q_sz^s}}{q_s}\right),
\ear
where
\beq{5.27n}
 \omega_0 = \sqrt{2{\cal E}_0}/q_0,
\qquad
 \omega_s = \sqrt{\frac14- 2 {\cal E}_s\nu_s^2},
\eeq
$B_\omega^0,B_\omega^s=I_\omega,K_\omega$ are the modified Bessel function.

The general solution of the WDW equation (\ref{4.7n}) is a superposition
of the ``separated'' solutions (\ref{5.20n}):
\bear{5.28} \lal
 \Psi(z) =\sum_B\int dpdPd{\cal E} C(p,P,{\cal E},B)
  \Psi_*(z|p,P,{\cal E},B),\nnn
\ear
where $p=(p_a)$, $P=(P_s)$, ${\cal E}=({\cal E}_s)$, $B=(B^0,B^s)$,
$B^0,B^s=I,K$, and $\Psi_*=\Psi_*(z|p,P,{\cal E},B)$ is given by the
relations (\ref{5.20n}), (\ref{5.24n})--(\ref{5.27n}) with ${\cal E}_0$
>from (\ref{5.23n}). Here $C(p,P,{\cal E},B)$ are smooth enough functions.
For several intersecting $p$-branes (non-composite electric and composite
electro-magnetic) see \cite{GrIM} and \cite{IMJ}, respectively.

\section{Black hole (BH) solutions}

\subsection{Classical BH solutions}

Let us single out solutions with a horizon (with respect to time $t$).
We put
\beq{5.65}
 1 \in I_s,
\eeq
i.e., the $p$-brane contains the time manifold. Let
\beq{5.66}
  \varepsilon_s = -1
\eeq
This is a physical restriction satisfied when a pseudo-Euclidean brane in
pseudo-Euclidean space is considered.

We single out the solution with a horizon:
for integration constants we put $\bar{c}^A = 0$,
\bear{5.67}
 c^A \eql \bar{\mu} \sum_{r = 0, s} \frac{U^{rA}}{(U^r,U^r)}
        - \bar{\mu} \delta^A_1, \\                      \label{5.68}
 C_0 \eql C_s =  \bar{\mu}^2,
\ear
where $\bar{\mu} > 0$. Here $A = (i_A, \alpha_A)$ and $A =1$ means $i_A =
1$.  It may be verified that the restrictions (\ref{5.47n})--(\ref{5.49n})
and (\ref{5.55n}) are satisfied identically.

Let us introduce the new radial variable $R = R(u)$ by the relations
\beq{5.69}         \nq
 \e^{- 2\bar{\mu} u} = 1 - \frac{2\mu}{R^{\bar{d}}},  \quad
 \mu = \bar{\mu} \bar{d} >0, \quad  \bar{d} = d_0 -1
\eeq
and put  $u_0 = 0$, $u_s < 0$,
\beq{5.70}
 \frac{|Q_s|}{\bar{\mu} \nu_s} \sinh \beta_s =1, \qquad
  \beta_s \equiv \bar{\mu}| u_s|,
\eeq
$s =e, m$.

Then the  solutions for the metric and the scalar fields
(see (\ref{5.46n}), (\ref{5.63n})) are:
\bear{5.72n}  \lal
 g = H_s^{2  d(I_s)\nu_s^2/(D-2)}
   \biggl\{ \frac{dR \otimes dR}{1 - 2\mu / R^{\bar d}}
     + R^2  d \Omega^2_{d_0}  \nnn
- H_s^{-2  \nu_s^2}
 \biggl(1 - \frac{2\mu}{R^{\bar d }} \biggr)  dt \otimes dt
+ \sum_{i = 2}^{n}   H_s^{-2 \nu_s^2 \delta_{i I_s}} g^i  \biggr\},
 \\ \lal
\label{5.73} \varphi^\alpha= \nu_s^2 \chi_s \lambda_{a_s}^\alpha \ln H_s,
\ear
where
\beq{5.74}
 H_s = 1 + \frac{{\cal P}_s}{R^{\bar{d}}},
 \qquad {\cal P}_s \equiv \frac{|Q_s|\bar{d}}{\nu_s} e^{- \beta_s}.
\eeq

The form field is given by  (\ref{2.28n}), (\ref{2.29n}) with
\bear{5.75} \lal
 \Phi^s = \frac{\nu_s}{H_s^{'} },  \\     \label{5.76}\lal
 H_s^{'}=
1 + \frac{{\cal P}_s^{'}}{ R^{\bar{d}} + {\cal P}_s - {\cal P}_s^{'} }, \\
\label{5.77}\lal
 {\cal P}_s^{'} \equiv - \frac{Q_s\bar{d}}{\nu_s}.
\ear
$s = e,m$. It follows from (\ref{5.70}), (\ref{5.74}) and (\ref{5.77})
that
\beq{5.78}
|{\cal P}_s^{'}| = \frac{\mu}{\sinh \beta_s} = {\cal P}_s e^{\beta_s}
= \sqrt{{\cal P}_s ({\cal P}_s +2\mu)}.
\eeq

The Hawking ``temperature" corresponding to
the solution is (see also \cite{BIM,O})
\beq{5.79}
 T_H = \frac{\bar{d}}{4 \pi (2 \mu)^{1/\bar{d}}}
 \left(\frac{2 \mu}{2 \mu + {\cal P}_s}\right)^{\nu_s^2}.
\eeq
Recall that $\nu_s^2 = (U^s,U^s)^{-1}$.

\medskip\noi
{\bf Extremal case.} In the extremal case $\mu \to +0$
we get for the metric (\ref{5.72n})
\bear{5.72ne} \lal
 g= H_s^{2  d(I_s)\nu_s^2/(D-2)}
 \biggl\{ dR \otimes dR + R^2  d \Omega^2_{d_0}  \nnn
- H_s^{-2  \nu_s^2}   dt \otimes dt
    + \sum_{i = 2}^{n}   H_s^{-2 \nu_s^2 \delta_{i I_s}} g^i  \biggr\},
\ear
where  $H_s = H_s^{'}$ (${\cal P}_s^{'} ={\cal P}_s$)
in (\ref{5.75}), $s =e,m.$

\medskip\noi
{\bf Remark.} This solution has a regular horizon at $R \to +0$
(with a finite limit of the Riemann tensor squared for $R \to +0$) if
$\nu_s^2 d(I_s) \bar{d} \geq D-2$  (see \cite{IMmp}).

\subsection{Quantum analogues of BH solutions}

Let us put $a =0$ in the WDW equation. In this case there exist a map that
puts into correspondence some quantum solution to WDW equation to any
classical BH solution.  We also put
\bear{5.80}
 {\cal E}_0 = E_0, \qquad  {\cal E}_s  = E_s,
\ear
i.e., the classical energies of subsystems coincide with the
eigenvalues of the Hamiltonians $\hat H_0$ and $\hat H_s$, respectively,
and
\bear{5.81}
 P_s = Q_s = - {\cal P}_s^{'} \frac{\nu_s}{\bar{d}},
\ear
i.e., the classical charges coincide with the eigenvalues
of the momentum operators $\hat P_s = - \p/\p \Phi^s$.

Then it may be shown that
\bear{5.82}
p_a z^a = c_A x^A.
\ear
Using these relations, we get solutions to the WDW equation (with $a =0$)
of special type. These solutions correspond to classical BH solutions and
have an ambiguity in the choice of the Bessel functions. We note that the
quantum energy constraint (\ref{5.23n}) is satisfied identically due to our
choice $a = 0$.

\medskip\noi
{\bf Extremal case.}  In the extremal case we have
\bear{5.83}
 {\cal E}_0 = {\cal E}_s  = p_a z^a= 0.
\ear
Hence the wave function (\ref{5.20ne}) reads
\beq{5.20ne}
 \Psi_*= \Psi_0 \Psi_s \exp(\im Q_s),
\eeq
where
\bear{5.24ne}
 \Psi_0 \eql B_{0}^0 \left( \im |\theta|  r_0 v^0 \right),
\\ \label{5.25ne}
 \Psi_s \eql  v_s^{1/2} B_{1/2}^s\left( \im Q_s v^s\right),
\ear
$r_0 = \sqrt{d_0(d_0 -1)}$ and
\bear{5.24nen}
 v_0 \eql \e^{q_0z^0} =  \exp(-\phi^0+ \sum_{i=0}^n d_i\phi^i),
\\ \label{5.25nen}
 v_s \eql \e^{q_sz^s} =
  \exp(-\chi_s\lambda(\varphi) + \sum_{i\in I_s}d_i\phi^i),
\ear
are ``quasivolumes", $s = e,m$.  The gravitational part of the wave
function, i.e. $\Psi_0$, coinsides with that of Refs.\,\cite{Zh1,Zh2},
see also \cite{BIMZ, IM3}.  For small quasivolumes $v_s \to 0$ we get
\bear{5.24ab}
 \Psi_s \lal\sim  v^s \sqrt{2 i Q_s/ \pi }, \qquad B =I,
\\ \label{5.25ab}
 \Psi_s \lal\sim \sqrt {\pi /2 i Q_s}, \qquad B =K.
\ear

For a big brane quasivolume $v_s \to \infty$  we get
\bear{5.24ba}
\Psi_s \lal\sim  \frac{\exp(iQ_s v^s)}{\sqrt{2 \pi i Q_s}}, \qquad B =I,
\\ \label{5.25ba}
\Psi_s \lal\sim  \frac{\exp(-iQ_s v^s)}{\sqrt{2 i Q_s/ \pi}}, \qquad B =K,
\ear
Thus, for a positive charge  ${\cal P}_s$ or, equivalently, $Q_s < 0$
(see (\ref{5.81})), the brane part of the solution with $B=K$ satisfies
the outgoing-wave boundary condition, used first in quantum cosmology
in \cite{25}, and is regular for small brane quasi-volumes.

\subsubsection{Example: $D =11$ supergravity}

Consider $D =11$ supergravity \cite{CJS} with the
truncated bosonic action (without a Chern-Simons term)
\beq{1.s}
S_{11tr} =\int_{M} d^{11}z \sqrt{|g|}
   \biggl\{ {R}[g] - \frac{1}{4!} F^2 \biggr\},
\eeq
where  $F$ is a $4$-form. Here we have two types of solutions:
an electric $2$-brane with $d(I_s) = 3$ ($s =e$) and
a magnetic $5$-brane with $d(I_s) = 6$ ($s = m$). In both cases
$(U^s,U^s) = 2 = \nu_s^{-2}$, ($s = e,m$. We put
$\eps(1) = -1$ and $\eps(k) = 1$, $k > 1$.

In the extremal case we get for an $M2$-brane ($s =e$) and an $M5$-brane
($s =m$):
\bear{2.s}      \lal
\Psi_{*} = B_{0}^0  \left( \im |\theta|  r_0 v^0 \right)v_s^{1/2}
   \times \nnn \times
 B_{1/2}^s \left( - \im {\cal P}_s \frac{1}{2 \bar{d} } v^s\right)
 \exp\biggl(- \im {\cal P}_s \frac{\nu_s}{\bar{d}} \Phi^s\biggr),
\ear
with $v_s = \exp( \sum_{i\in I_s}d_i\phi^i)$, $s = e,m$, and
$v_0$  defined in (\ref{5.24nen}).

\subsubsection{Example: $D =4$ Einstein-Maxwell gravity}

Consider $D =4$ gravity:
\beq{1m.s}
 S_{4} = \int_{M} d^{4}z \sqrt{|g|}
  \biggl\{ {R}[g] - \frac{1}{2!} F^2 \biggr\},
\eeq
where  $F$ is a $2$-form. Here we have two types of solutions:
an electric $0$-brane (electric charge) with $d(I_s) = 1$ ($s =e$) and
a magnetic $0$-brane (magnetic charge) with $d(I_s) = 1$ ($s = m$). In both
cases $(U^s,U^s) = 1/2 = \nu_s^{-2}$ ($s = e,m$). We put $\eps(1) = -1$.
Here $n = 1$, $d_0 =2$ and $d_1 =1$.

The (extremal) solutions read
\bear{2m.s}                  \lal
\Psi_{*} = B_{0}^0  \left( \im  |\theta|  \sqrt{2} v^0 \right) v_s^{1/2}
  \times \nnn \times
\quad B_{1/2}^s  \left( - \im {\cal P}_s  v^s \right)
   \exp(- \im {\cal P}_s \nu_s \Phi^s),
\ear
with $v_s = \exp(\phi^1)$, $s = e,m$, and $v_0 = \exp( \phi^0+ \phi^1)$. We
see that here electric and magnetic solutions coincide.  (This fact
may be concidered as a simple manifestation of electro-magnetic duality at
a quantum level).

\subsubsection{WDW equation with fixed charges}

There exists another quantization scheme, where the fields of forms are
considered to be classical. This scheme is based on the zero-energy
constraint relation (\ref{5.33n}), see \cite{LMMP}.  The corresponding WDW
equation in the harmonic gauge reads
\beq{5.80a}\nq
 \hat{H}_Q \Psi \equiv
 \left(-\frac{1}{2\theta}\ {\bar G}^{AB}
 \frac{\d}{\d x^A} \frac{\d}{\d x^B}+\theta V_Q \right) \Psi=0
\eeq
where the potential $V_Q$ is defined in (\ref{5.32n}). This equation
describes quantum cosmology with classical fields of forms
and quantum scale factors and dilatonic fields.

The basis of solutions is given by the following replacements
in (\ref{5.20n}), (\ref{5.23n}), (\ref{5.25n}) and  (\ref{5.27n}):
$P_s  \mapsto Q_s$, $2aR[{\cal G}] \mapsto 0$,
$\omega_s \mapsto \sqrt{- 2{\cal E}_s \nu_s^2}$.
and $\Psi_s(z^s)  \mapsto B_{\omega_s}^s
\left(\sqrt{\eps_s Q_s^2}\e^{q_sz^s}/q_s\right)$.

In this approach there is no problem with the $a$ parameter when quantum
analogues of black-hole solutions are constructed (since in the harmonic
gauge $R({\bar G}) =0$).

Let us compare quantum black-hole solutions in the two approaches.
The function $\Psi_0$ is the same in both
approaches but for the brane part of the wave function we get
\bear{5.25pope}
 \Psi_s= B_{0}^s \left( \im Q_s v^s\right).
\ear
For small  quasivolumes $v_s \to 0$ we get
\bear{5.24amer}          \lal
 \Psi_s \sim  1, \inch  B =I,    \\ \label{5.25amer}\lal
 \Psi_s \sim - \ln( i Q_s v^s/2), \qquad B =K.
\ear

For a big brane quasivolume, $v_s \to \infty$,  we get
\bear{5.24as}  \lal
\Psi_s \sim  \frac{\exp(iQ_s v^s)}{\sqrt{2 \pi i Q_s v^s}}, \qquad B =I,
\\ \label{5.25as}  \lal
\Psi_s \sim  \frac{\exp(-iQ_s v^s)}{\sqrt{2 i Q_s v^s/ \pi}}, \qquad B =K,
\ear

Thus for $Q_s < 0$  the brane part of the solution with $B=K$ satisfies the
outgoing-wave boundary condition \cite{25} but it is not regular for a
small brane quasi-volume.

\section{Conclusions}

We have considered classical spherically symmetric  solutions with one
brane and the corresponding black-hole solutions.  Using solutions to the
Wheeler-DeWitt equation, have we suggested the quantum analogues to
black-hole solutions in the extremal and non-extremal cases. This was
possible when the coupling parameter in the WDW equation was trivial:  $a
=0$. In the alternative approach of \cite{LMMP} (with classical fields of
forms) one may use an arbitrary coupling $a$ in the WDW equation when
constructing quantum analogues.

\Acknow
{This work was supported in part by a DFG grant, by the Russian
Ministry for Science and Technology and Russian Basic Research Foundation,
grant N 98-02-16414. V.D.I. and V.N.M.  are grateful for hospitality
during their stay at Nara Women's University.}

\end{document}